\documentclass[english]{article}
\usepackage[T1]{fontenc}
\usepackage[latin9]{inputenc}
\usepackage{amsmath}
\usepackage{graphicx}
\usepackage{esint}

\newcommand{\lyxrightaddress}[1]{
\par {\raggedleft \begin{tabular}{l}\ignorespaces
#1
\end{tabular}
\vspace{1.4em}
\par}
}

\usepackage{babel}

\begin{document}

\title{The Maxwell-Boltzmann Distribution is not the Equilibrium on a Hyperboloid }

\author{S. G. Rajeev}

\maketitle

\lyxrightaddress{Department of Physics and Astronomy, University of Rochester, Rochester
NY 14618.}

We give a geometric formulation of the Fokker-Planck-Kramer equations
for a particle moving on a Lie algebra under the influence of a dissipative
and a random force. Special cases of interest are fluid mechanics,
the Stochastic Loewner Equation and the rigid body. We find that the
Boltzmann distribution, although a static solution, is not normalizable
when the algebra is not unimodular. This is because the invariant
measure of integration in momentum space is not the standard one.
We solve the special case of the upper half-plane (hyperboloid) explicitly:
there is another equilibrium solution to the Fokker-Planck equation,
which is integrable. It breaks rotation invariance; moreover, the
most likely value for velocity  is not zero.

\section{Introduction}

Arnold \cite{Arnold} showed that the Euler equations of an ideal
fluid describe geodesics in the group of volume preserving co-ordinate
transformations of the space filled by the fluid. The negative curvature
of this geometry implies that the geodesics diverge from each other
exponentially: an elegant explanation for the observed instability
of fluid motion. A more realistic description would include dissipation
(Navier-Stokes) as well as small random forces not included in the
ideal model. Since the infinite dimensional fluid system is difficult
to understand directly, we look for analogous finite dimensional examples
that have (i) Negative Curvature (exponentially diverging trajectories),
(ii) Dissipation and (iii) Random Forcing.

The celebrated Langevin equation for the velocity of a particle in
Brownian motion includes (2) and (3); being one dimensional, it cannot
have curvature.

\[
\frac{dv}{dt}=-\gamma v+\eta.\]

Here $\gamma$ is the dissipation constant, proportional to the viscosity
of the medium in which the particle is moving. Also, $\eta$ is a
random force, modelled as a Gaussian of zero mean and zero correlation
time({}``white noise''):

\[
<\eta(t)\eta(t')>=2D\delta(t-t').\]

The constant $D$ measures the strength of the fluctuations. This
leads to the Fokker-Planck equation for the probability density of
velocity

\[
\frac{\partial P}{\partial t}=\frac{\partial}{\partial v}\left[D\frac{\partial P}{\partial v}+\gamma vP.\right]\]

The equilibrium (static) solution of this equation is the Boltzmann
distribution

\[
P(v)=\frac{1}{Z}e^{-\beta\frac{v^{2}}{2}}\]

with an inverse temperature given by the Einstein relation

\[
\beta D=\gamma.\]

How would things change if the particle is moving in a space of negative
curvature? We will study in this paper the simplest such model. We
will show, by explicit solution, that the equilibrium solution is
not always the Boltzmann distribution. In particular, the most likely
value of velocity is not zero. The essential reason is that the invariant
measure on the space of velocities is not translation invariant: the
Boltzmann distribution is not integrable in this measure. There is
another static solution to the Fokker-Planck which is integrable.
In simple cases we can find it explicitly in terms of classical functions.

There is already an extensive literature on Brownian motion on curved
manifolds. The main approach is to solve the heat equation in position
space. However, this does not address the question of the equilibrium
velocity distribution, which we study here.

\section{The Fokker-Planck-Kramers Equation}

We review theory of a dynamical system subject to dissipation and
fluctuation, paying special attention to the role of the measure on
phase space in the Fokker-Plank-Kramers equation. We base our discussion
on refs. \cite{ChandraStochastic,Kramers} but use a more geometric
formulation. The erudite reader might want to skip this section after
a glance at the notation.

\subsection{The Ideal System}

Consider a dynamical system with equations of motion

\[
\frac{du_{a}}{dt}=V_{a}\]

for some vector field $V^{a}$. We will assume that there is a measure
$\mu$ that is invariant under this evolution,

\begin{equation}
\partial^{a}\left[\mu V_{a}\right]=0;\label{eq:InvariantMeasure}\end{equation}

as well as a quantity $E$(energy) that is conserved by it:

\[
\frac{dE}{dt}=V_{a}\frac{\partial E}{\partial u_{a}}=0.\]

For example, if there is a sympletic form $\omega_{ab}$ and hamiltonian
$H$ such that \[
V_{a}=\left\{ H,u_{a}\right\} \]

for the induced Poisson bracket, the invariant measure is the Pfaffian
(square root of the determinant) of $\omega:$

\[
\mu=\mathrm{Pf}\ \omega.\]

Then (\ref{eq:InvariantMeasure}) is just Liouville's theorem. Energy
will be conserved if

\[
\left\{ H,E\right\} =0\]
 This is automatic if the hamiltonian itself is the energy; but this
is not necessary.

\subsection{Dissipation}

A more realistic description could include a dissipative force. We
will assume that the dissipative force is a gradient of energy; i.e.,

\[
\Gamma_{a}=-\Gamma_{ab}\frac{\partial E}{\partial u_{b}}\]

for some positive%
\footnote{That is, $\Gamma_{ab}$is symmetric and $\Gamma_{ab}u^{a}u^{b}\geq0$
for all $u_{a}.$%
}$\Gamma_{ab}.$ The resulting equations

\[
\frac{du_{a}}{dt}=\Gamma_{a}+V_{a}\]

imply that energy decreases:

\[
\frac{dE}{dt}=-\Gamma_{ab}\frac{\partial E}{\partial u_{a}}\frac{\partial E}{\partial u_{b}}\leq0.\]

\subsection{Fluctuation}

We assume a standard model of fluctuating force: a Gaussian $\eta_{a}(t)$
with zero correlation time ( {}``white noise'')

\[
<\eta_{a}(t)\eta_{b}(t')>=2D_{ab}\delta(t-t').\]

The correlation tensor $D^{ab}$must be positive as well. The time
evolution is now a system of stochastic ordinary differential equations

\[
\frac{du_{a}}{dt}=\Gamma_{a}+V_{a}+\eta_{a}.\]

The probability density $P(u,t)$ for the random variable $u$ will
then satisfy a partial differential equation: a diffusion equation
with drift. The total probability

\[
\int P\mu du=1\]

so that $\frac{\partial[\mu P]}{\partial t}$ must be a total derivative.
This leads us to the Fokker-Planck-Kramers equation

\[
\mu\frac{\partial P}{\partial t}=\frac{\partial}{\partial u_{a}}\left[\mu\left\{ D_{ab}\frac{\partial P}{\partial u_{b}}-(\Gamma_{a}+V_{a})P\right\} \right].\]

\subsection{The Boltzmann Solution}

The function

\begin{equation}
P=e^{-\beta E}\label{eq:Boltzmann}\end{equation}

is a solution to this equation provided that the dissipation and fluctuation
tensors are proportional (the Einstein-Smoluchowsky relation):

\begin{equation}
\beta D_{ab}=\Gamma_{ab}.\label{eq:FluctuationDissipationTheorem}\end{equation}

For, in this case

\[
D_{ab}\frac{\partial P}{\partial u_{b}}=\Gamma_{a}P.\]

Moreover,

\[
\frac{\partial\left[\mu V^{a}P\right]}{\partial u_{a}}=\frac{\partial\left[\mu V^{a}\right]}{\partial u_{a}}P+\mu V_{a}\frac{\partial P}{\partial u_{a}}=\mu\{H,P\}=-\mu\beta E\left\{ H,E\right\} =0\]

since the ideal evolution $V^{a}$ preserves the measure $\mu$ and
conserves energy.

If, in addition to (\ref{eq:FluctuationDissipationTheorem}) the convergence
condition

\begin{equation}
Z=\int e^{-\beta E}\mu du<\infty\label{eq:Convergence}\end{equation}
is satisfied, we can normalize the solution and get equilibrium probability
distribution

\[
\frac{1}{Z}e^{-\beta E}.\]

If (\ref{eq:FluctuationDissipationTheorem}) or (\ref{eq:Convergence})
is violated, the equilibrium is not given by the Boltzmann distribution.
We will exhibit an example below where (\ref{eq:FluctuationDissipationTheorem})
holds but not ( \ref{eq:Convergence}). In our case, there is another
static, normalizable solution which we will determine explicitly.

In general, there may be no equilibrium state; or the system might
reach a steady state which dissipates energy at some constant rate.

\subsection{The Adjoint Equation}

Suppose that the Einstein relation holds, so that $e^{-\beta E}$is
a static solution. If we make the change of variables

\[
P=e^{-\beta E}Q\]

the FPK equation becomes

\begin{equation}
\frac{\partial Q}{\partial t}=\frac{1}{\mu}\frac{\partial}{\partial u_{a}}\left[\mu D_{ab}\frac{\partial Q}{\partial u_{b}}\right]+\left[\Gamma_{a}-V_{a}\right]\frac{\partial Q}{\partial u_{a}}\label{eq:AdjointFKP}\end{equation}

Even when the Boltzmann solution is not normalizable, this is a good
starting point to search for a normalizable static solution.

\subsection{Fast Dynamics}

If the dynamics $V^{a}$ is much faster than the dissipation and fluctuation
effects, the details of the vector field do not matter: the system
will wander around in phase space and fill it. A vestige of the ideal
dynamics survives: the invariant measure $\mu.$ (A kind of micro-canonical
ensemble.) In this limit we get the Fokker-Planck equation

\[
\mu\frac{\partial P}{\partial t}=\partial_{a}\left[\mu\left\{ D^{ab}\partial_{b}P-\Gamma^{a}P\right\} \right].\]

Again, if the Einstein relation holds,we have the adjoint equation,

\[
\frac{\partial Q}{\partial t}=\frac{1}{\mu}\frac{\partial}{\partial u_{a}}\left[\mu D_{ab}\frac{\partial Q}{\partial u_{b}}\right]+\Gamma_{a}\frac{\partial Q}{\partial u_{a}}.\]

\section{Stochastic Geodesic Motion on Groups}

There are several examples in physics where we are interested in geodesic
motion on Lie groups. Arnold's observation that the Euler equations
of an ideal fluid are geodesic equations on the group of incompressible
diffeomorphisms is the most important example. Navier-Stokes equations
follow from adding a dissipation and many standard discussions of
turbulence involve adding a Gaussian random force\cite{Kraichnan}
. Self-erasing random walks can reformulated as random walks on the
diffeomorphism group of the circle, the Schramm-Loewner Equation\cite{SLE}.
This dynamics takes place on a generalization of the upper half-plane,
the space of univalent functions. A much more elementary example would
be rigid body motion; random motion of rigid bodies have been used
to model dust grains in astronomy\cite{RigidBody}. We will now give
a general framework for this theory,. The ingredients are the structure
constants $f_{ab}^{c}$ of the Lie algebra and three symmetric positive
tensors $G^{ab},\Gamma^{ab},D^{ab}$characterizing the energy, dissipation
and fluctuations respectively. The main result is that the Boltzmann
distribution is the equilibrium distribution only for unimodular Lie
algebras: the trace of the matrices in the adjoint representation
must vanish.

We will study in detail the simplest example that is not unimodular.
There is an equilibrium distribution, but it is not the Boltzmann
distribution.

\subsection{Geodesic Motion as Hamiltonian Dynamics}

Geodesics on a Riemann manifold form a hamiltonian system \cite{Chavel}
. If the manifold is a Lie group, and the metric is invariant under
the left action of the group on itself, the evolution of the tangent
vector ({}``momentum'') is independent of the position and can be
studied separately\cite{Milnor,hydrochennai}. The Poisson Brackets
of velocity components are just the commutation relations of the Lie
algebra in some basis:

\[
\left\{ v_{a},v_{b}\right\} =f_{ab}^{c}v_{c}.\]

A left-invariant metric on the group is the same thing as a quadratic
form on the Lie algebra ({}``:energy'')

\[
E=\frac{1}{2}G^{ab}v_{a}v_{b}.\]
The geodesic equations follow by the usual rules of hamiltonian mechanics:

\[
\frac{dv_{a}}{dt}=\left\{ v_{a},E\right\} =f_{ab}^{c}G^{bd}v_{c}v_{d}\equiv V_{a}.\]

In the special case of the rigid body, the Lie algebra is $SO(3)$;
the tensor $G^{ab}$ is the inverse of moment of inertia and the geodesic
equations are the Euler equations of the rigid body. If the metric
on the group is also right invariant, the tensor $G^{ab}$ is isotropic

\[
f_{ab}^{c}G^{bd}+f_{ab}^{d}G^{bc}=0.\]

and the momentum is a constant. This is true of the isotropic rigid
body (equal moment of inertia in all directions) but is not usually
the interesting case.

More generally, a function $H(\rho)$ where $\rho^{2}=G^{ab}v_{a}v_{b}$
can be chosen as the hamiltonian. The equations of motion become

\[
\frac{dv_{a}}{dt}=\left\{ v_{a},H\right\} =\frac{H'(\rho)}{\rho}f_{ab}^{c}G^{bd}v_{c}v_{d}\equiv V_{a}\]

The solutions are still geodesics, differing only by a constant reparametrization:
the time variable gets mutliplied by a constant (which can depend
on energy). We will see that an unconventional choice $H(\rho)=-\frac{k}{\rho}$
will simplify the FPK equation in the special case we study in detail
below.

\subsubsection{Invariant Measure}

The obvious measure of integration $dv=dv_{1}\wedge dv_{2}\cdots$
is only invariant if

\[
\frac{\partial V_{a}}{\partial v_{a}}=0.\]

For a Hamiltonian $H,$

\[
V_{a}=\left\{ v_{a},H\right\} =f_{ab}^{c}v_{c}\frac{\partial H}{\partial v_{b}}\]

so that

\[
\frac{\partial V_{a}}{\partial v_{a}}=f_{ab}^{a}\frac{\partial H}{\partial v_{b}}.\]

A Lie algebra is said to be unimodular if the trace of the structure
constants is zero:

\[
f_{ab}^{a}=0.\]

This is the condition for the measure $dv$ on the Lie algebra to
be invariant under the adjoint action. The Haar measure on the corresponding
Lie group will be both left and right invariant. Any semi-simple Lie
algebra is unimodular; such as the rotations $SO(n),$linear canonical
transformations $Sp(n),$unitary transformations $U(n)$ or any products
of these groups.

An example that is not unimodular is the Lie algebra of the affine
group: the only non-abelian Lie algebra in two dimensions. Typical
examples are solvable and nilpotent algebras or Lie algebras containing
them as subalgebras. When the algebra is not unimodular, there is
still a measure $\mu dv$ that is invariant under the adjoint action:
it is not anymore just $dv.$

\subsection{Dissipation}

Adding a fluctuation that is a gradient of energy leads to an Ohmic
force

\[
\Gamma_{a}=-\Gamma_{ad}G^{db}v_{b}\equiv\Gamma_{a}^{b}v_{b}\]

and

\[
\frac{dv_{a}}{dt}=-\Gamma_{a}^{b}v_{b}+f_{ab}^{c}G^{bd}v_{c}v_{d}.\]

\subsubsection{The Navier-Stokes Equation}

A special case is the Navier-Stokes equation. The divergence free
vector fields form a Lie algebra under the usual commutator. The energy
is just the $L^{2}$-norm

\[
E=\frac{1}{2}\int v_{i}v_{i}dx\equiv\frac{1}{2}G^{ab}v_{a}v_{b}.\]

The tensor $G^{ab}$ is just the Dirac delta function.

The dissipation tensor, thought of as a quadratic form on vector fields
is the $H_{1}$norm:

\[
\Gamma^{ab}v_{a}v_{b}\equiv\int\partial_{i}v_{j}\partial_{i}v_{j}dx\]
 This leads (after some calculations \cite{hydrochennai}) to the
Navier-Stokes equations

\[
\frac{\partial v_{i}}{\partial t}=\partial^{2}v_{i}-v_{j}\partial_{j}v_{i}-\partial_{i}p\]

where the pressure $p$is determined from the constraint

\[
\partial_{i}v_{i}=0.\]

Replacing the space of incompressible vector fields by a finite dimensional
Lie algebra allows to study simpler models of this important physical
system.

\subsection{Fluctuation}

Adding a random force leads to a Langevin equation

\[
\frac{dv_{a}}{dt}=-\Gamma_{a}^{b}v_{b}+f_{ab}^{c}G_{cd}v^{b}v^{d}+\eta_{a}.\]

The Fokker-Planck-Kramers equation becomes

\[
\frac{\partial P}{\partial t}=\frac{1}{\mu}\frac{\partial}{\partial u_{a}}\left[\mu\left\{ D_{ab}\frac{\partial P}{\partial u_{b}}+\left(\Gamma_{a}^{b}v_{b}-f_{ab}^{c}G_{cd}v^{b}v^{d}\right)P\right\} \right]\]

There is a static solution

\[
P_{B}(v)=e^{-\beta\frac{v^{2}}{2}}\]

if

\[
\beta D_{ab}=\Gamma_{ab}.\]

This can be interpreted as a probability distribution if

\[
\int\mu e^{-\beta\frac{v^{2}}{2}}dv\]

converges.

\subsubsection{The Randomly Forced Navier-Stokes Equation}

In fluid mechanics, the Navier-Stokes equation with random forcing

\[
\frac{\partial v_{i}}{\partial t}=\partial^{2}v_{i}-v_{j}\partial_{j}v_{i}-\partial_{i}p+\eta_{i}(x,t)\]

is often used to model turbulence. The correlation tensor of fluctuation
is chosen to be translation invariant \[
D_{ij}(x,y)=\int\left[\delta_{ij}-\frac{k_{i}k_{j}}{k^{2}}\right]\tilde{D}(p)dp.\]

The function $\tilde{D}(p)=p^{2}$ would satisfy the condition (\ref{eq:FluctuationDissipationTheorem}).
But it is not the one usually used in the literature on turbulence.
Generally speaking, dissipation is important at short distance scales
(viscous force is proportional to the gradient of velocity) while
fluctuations are believed to pump energy into the system at large
distance scales. Thus $\tilde{D}(p)=|p|^{-\alpha}$ , with a negative
power of momentum is more reasonable\cite{Kraichnan}. So an equilibrium
of the Boltzmann type is unlikely to exist in fluid mechanics: instead,
a steady state solution in which energy is dissipated at a constant
rate is more likely to be the correct answer. The theory of Kolmogorov
that leads to scale invariant velocity correlations, in one such example.

\section{The Langevin Equation on a Hyperboloid}

It is often useful to replace an infinite dimensional physical system
by a finite dimensional toy model, which still has some of the basic
structure intact. In our case, it is important that the underlying
Lie algebra be non-abelian. Otherwise it simply reduces to the standard
Brownian motion in Euclidean space. Thus to get something non-trivial,
the Lie algebra must be at least two dimensional.

The only non-abelian Lie algebra in two dimensions is

\begin{equation}
\left\{ v_{0},v_{1}\right\} =v_{1}.\label{eq:PB}\end{equation}

The corresponding Lie group is the set of triangular matrices

\[
\left(\begin{array}{cc}
a_{0} & a_{1}\\
0 & 1\end{array}\right)\]

with $a_{0}>0.$ The upper half-plane

\[
\mathcal{U}=\left\{ (a_{0},a_{1})|a_{0}>0\right\} \]

parametrizes such matrices. The group multiplication law is

\[
(a_{0},a_{1})(a_{0}',a_{1}')=(a_{0}a_{0}',a_{0}a_{1}'+a_{1}).\]

This group is often called the affine group, as it acts on the real
line by translations and scaling (affine transformation)

\[
(a_{0},a_{1})=a_{0}t+a_{1}.\]

The natural geometry on the upper half plane is the Poincare' metric,
which has constant negative curvature: the simplest geometry with
negative curvature. There is a one-one correspondence of the hyperboloid
with the upper half-plane;the induced metric on the hyperboloid is
just the the Poincare' metric on the upper half-plane. A particle
constrained to move on the hyperboloid, but free of other forces,
will move along geodesics.

The Poincare' metric is invariant under the left action (but not the
right action) of the affine group. Because it has negative curvature,
the geodesics do not have constant tangent vectors: only the length
is preserved, while the direction changes through parallel transport.
It is straightfoward to derive the geodesic equations \cite{Milnor,hydrochennai}

\[
\frac{dv_{0}}{dt}=-v_{1}^{2}\]

\[
\frac{dv_{1}}{dt}=v_{0}v_{1}\]

where $t$ is the arc-length. Thee can be thought of as Hamilton's
equations implied by the Poisson brackets (\ref{eq:PB}) and the Hamiltonian

\[
E=\frac{1}{2}\left[v_{0}^{2}+v_{1}^{2}\right]\]

which is just the kinetic energy.

It is convenient to use a kind of hyperbolic analogue of the polar
co-ordinate system in velocity space

\begin{equation}
v_{0}=-\rho\tanh\theta,\ v_{1}=\rho\epsilon\mathrm{sech}\theta\label{eq:hyperbolicpolar}\end{equation}

in terms of which

\[
\frac{d\theta}{dt}=\rho\]

\[
\frac{d\rho}{dt}=0.\]

The {}``angular'' variable $\theta$ has the range $-\infty<\theta<\infty.$It
is canonically conjugate to the radial (or {}``action'') variable
$\rho$

\[
\left\{ \rho,\theta\right\} =1\]

under the above Poisson bracket (\ref{eq:PB}). The discrete variable
$\epsilon=\pm1$ is needed in addition to cover both signs of $v_{1}.$
The solutions are semi-circles in the $(v_{0},v_{1})$plane:

\[
v_{0}=-\rho\tanh\rho t\]

\[
v_{1}=\rho\epsilon\mathrm{sech}\ \rho t.\]

Such a geodesic motion on a space of negative curvature models the
Euler equation of an ideal fluid, which are geodesic equations on
the group of volume preserving diffeomorphisms. Indeed, the affine
Lie algebra is a subalgebra of the incompressible vector fields. In
a plane for example, the pair of vector fields

\[
-x\frac{\partial}{\partial x}+y\frac{\partial}{\partial y},\ y\frac{\partial}{\partial x}\]
span such a subalgebra.

\section{Dissipation}

We now modify the above system to include a dissipation, to model
the Navier-Stokes equation

\[
\frac{dv_{0}}{dt}=-\gamma v_{0}-v_{1}^{2}\]

\[
\frac{dv_{1}}{dt}=-\gamma v_{1}+v_{0}v_{1}\]

In terms of hyperbolic polar co-ordinates $\rho,\theta$ as above(\ref{eq:hyperbolicpolar}),

\[
\frac{d\rho}{dt}=-\gamma\rho,\quad\frac{d\theta}{dt}=\rho.\]

The energy $E(\rho)=\frac{1}{2}\rho^{2}$ is then monotonically decreasing:\[
\frac{dE}{dt}=-2\gamma E\]

But the equations can still be solved analytically. The solution that
passes through the point with hyperbolic polar co-ordinates $(\rho_{1},\theta_{1})$
at time $t_{1}$ is,

\[
v_{0}=-\rho_{1}e^{-\gamma(t-t_{1})}\tanh\left[\theta_{1}+\rho_{1}\frac{1-e^{-\gamma(t-t_{1})}}{\gamma}\right]\]

\[
v_{1}=\rho_{1}\epsilon_{1}e^{-\gamma(t-t_{1})}\mathrm{sech}\ \left[\theta_{1}+\rho_{1}\frac{1-e^{-\gamma(t-t_{1})}}{\gamma}\right]\]

As $t\to\infty,$ the velocities tend to zero; as $t\to-\infty,$
$v_{1}(t)\to0$ and $v_{0}(t)\to-\infty.$ For intermediate values
it roughly follows the semi-circle.

If we choose as hamiltonian $H(\rho)=-\frac{k}{\rho},$we would get
instead

\[
\frac{d\rho}{dt}=-\gamma\rho,\quad\frac{d\theta}{dt}=H'(\rho)=\frac{k}{\rho^{2}}\]

which is just as easily solvable.

\section{Random Forcing}

Adding a random force with

\[
\mathrm{<\eta_{a}(t)>=0,\ <\eta_{a}(t)\eta_{b}(t')>=2D\delta_{ab}\delta(t-t').}\]

we get the Langevin equation on a hyperboloid

\begin{equation}
\frac{dv_{0}}{dt}=-\gamma v_{0}-v_{1}^{2}+\eta_{1}\label{eq:Langevin}\end{equation}

\[
\frac{dv_{1}}{dt}=-\gamma v_{1}+v_{0}v_{1}+\eta_{2}\]

As noted earlier, the affine Lie algebra is not unimodular so the
invariant measure is not $dv.$ It is instead $d\rho d\theta=\frac{dv^{0}dv^{1}}{v^{1}}\equiv\mu dv_{0}dv_{1}$
. The vector field $V_{a}$ arising from a hamiltonan $H$ and Poisson
Brackets (\ref{eq:PB}) preserve this measure:

\[
V_{0}=-v_{1}\frac{\partial H}{\partial v_{1}},\ V_{1}=v_{1}\frac{\partial H}{\partial v_{0}}\Rightarrow\frac{\partial}{\partial v^{a}}\left[\mu V_{a}\right]=0.\]

The above (\ref{eq:Langevin}) corresponds to H=$\frac{1}{2}v^{2}.$
The choice $H=\frac{k}{v}$ yields equivalent geodesic equations,
and is more convenient for solving the FKP equation.

We are led to the Fokker-Plank equation

\[
\frac{\partial P}{\partial t}=\frac{1}{\mu}\partial_{a}\left[\mu\left\{ D\partial_{a}P+(\gamma v_{a}-V_{a})P\right\} \right].\]

The function $e^{-\beta\frac{v^{2}}{2}}$ is a static solution if

\[
\beta D=\gamma.\]

But the integral

\[
\int e^{-\beta\frac{v_{0}^{2}+v_{1}^{2}}{2}}\frac{d^{2}v}{v^{1}}\]

is logarithmically divergent near $v^{1}=0$. In canonical co-ordinates
the measure is just the Liouville measure

\[
\mu d^{2}v=d\rho d\theta\]

In this language, we have a linear divergence in the angular co-ordinate:

\[
\int e^{-\frac{1}{2}\beta\rho^{2}}d\rho\int d\theta\]

Thus the Boltzmann distribution is not the correct equilibrium solution.

\subsection{The Adjoint Equation}

After the change of variables

\[
P=e^{-\beta\frac{v^{2}}{2}}Q\]

the FPK equation becomes its adjoint

\begin{equation}
\frac{\partial Q}{\partial t}=\frac{D}{\mu}\frac{\partial}{\partial v_{a}}\left[\mu\frac{\partial Q}{\partial v_{b}}\right]+\left[-\gamma v_{a}-V_{a}\right]\frac{\partial Q}{\partial v_{a}}\label{eq:AdjointHyperboloid}\end{equation}

Notice that the the first term on the r.h.s. is a kind of Laplacian,
but is \emph{not} the Lapalce-Beltrami operator: the symplectic volume
element implied by the Poisson brackets appears in place of the Riemannian
volume element . In polar co-ordinates

\[
\frac{\partial Q}{\partial t}=D\frac{\partial^{2}Q}{\partial\rho^{2}}-\gamma\rho\frac{\partial Q}{\partial\rho}+\frac{D}{\rho^{2}}\frac{\partial}{\partial\theta}\left[\cosh^{2}\theta\frac{\partial Q}{\partial\theta}\right]-H'(\rho)\frac{\partial Q}{\partial\theta}\]

We want a solution for which \[
\int_{0}^{\infty}d\rho e^{-\frac{\beta}{2}\rho^{2}}\int_{-\infty}^{-\infty}d\theta Q(\rho,\theta,t)\]

converges. In particular, it cannot be a constant in the $\theta$variable.

If the hamiltonian is $H=-\frac{k}{\rho}$ the last term would be
$\frac{k}{\rho^{2}}\frac{\partial Q}{\partial\theta}$ and the equation
would be solvable by separation of variables:

\[
Q(\rho,\theta,t)=Q_{1}(\theta)Q_{2}(\rho)e^{-D\beta_{2}t}\]

\[
\frac{\partial}{\partial\theta}\left[\cosh^{2}\theta\frac{\partial Q_{1}}{\partial\theta}\right]+a\frac{\partial Q_{1}}{\partial\theta}+\beta_{1}Q_{1}=0\]

\[
\frac{\partial^{2}Q_{2}}{\partial\rho^{2}}-\beta\rho\frac{\partial Q_{2}}{\partial\rho}-\frac{\beta_{1}}{\rho^{2}}Q_{2}+\beta_{2}Q_{2}=0\]

where $a=\frac{k}{D}.$ The separation constants $\beta_{1},\beta_{2}$
are eigenvalues of these differential operators; they depend on $a$
as well as on discrete `quantum numbers' labelling eigenfunctions.
For a static solution $\beta_{2}=0.$

\subsection{Angular Equation}

Change to

\[
u=\tanh\theta\]

\[
\frac{\partial^{2}Q_{1}}{\partial u^{2}}+2a\frac{\partial Q_{1}}{\partial u}+\frac{\beta_{1}}{1-u^{2}}Q_{1}=0\]

\[
Q_{1}=e^{-au}\sqrt{1-u^{2}}\phi(u)\]

\[
(1-u^{2})\phi''-2u\phi'+\left[\beta_{1}-a^{2}(1-u^{2})-\frac{1}{1-u^{2}}\right]\phi=0\]

This is the Prolate Angular Spheroidal Wave Equation\cite{AbrSt}.

A simple solution is $\phi(u)=\sqrt{1-u^{2}}$ when $a=0,\beta_{1}=2.$Then
$Q_{1}[u]=(1-u^{2})=\mathrm{sech}^{2}\ \theta$ is integrable. This
belongs to a different branch from the non-integrable constant solution,
for $a=0.$ When $a\neq0,$this solution continues to a solution $S_{11}(\theta)$integrable
in $\theta$, with $\beta_{1}>2.$

\subsection{Radial Wave Function}

For a static solution,

\[
\frac{\partial^{2}Q_{2}}{\partial\rho^{2}}-\beta\rho\frac{\partial Q_{2}}{\partial\rho}-\frac{\beta_{1}}{\rho^{2}}Q_{2}=0.\]

The solution is a confluent hypergeometric function. The solution
is

\[
Q_{2}=\rho^{\frac{1+\sqrt{1+4\beta_{1}}}{2}}F_{1}^{1}\left(\frac{1+\sqrt{1+4\beta_{1}}}{4},1+\frac{1}{2}\sqrt{1+4\beta_{1}},\frac{\beta r^{2}}{2}\right)\]

If $\beta_{1}=0,$a solution is constant, which is not normalizable
in the angular co-ordinate.

Now suppose $a=0,\beta_{1}=2,$for which the angular integral converges.
Then there is a solution that is normalizable in the radial variable
as well.

\[
Q_{2}=1-\frac{1}{\sqrt{\beta}\rho}e^{-\frac{\beta\rho^{2}}{4}}\mathrm{erf}\left(\frac{\sqrt{\beta}\rho}{2}\right)\]

Thus, although the Boltzmann distribution is not a normalizable solution,
there is another solution that is static and normalizable. But it
breaks rotation invariance spontaneously!. This solution can be continued
to $\beta_{1}>2$, and remains integrable, when the hamiltonian is
not zero. Thus rotation invariance is spontaneously broken and moreover,
the most likely value of energy is not zero.

\subsection{Equilibrium Solution}

We have putting all of the above together, in the limit of $k=0$
when the hamiltonian has a small effect,

\[
P(\rho,\theta)d\rho d\theta=e^{-\frac{\beta}{2}\rho^{2}}\left[1-\frac{1}{\sqrt{\beta}\rho}e^{-\frac{\beta\rho^{2}}{4}}\mathrm{erf}\left(\frac{\sqrt{\beta}\rho}{2}\right)\right]\mathrm{sech}^{2}\theta d\rho d\theta\]

or in the original co-ordinates

\[
P(v_{0},v_{1})\frac{dv_{0}dv_{1}}{v_{1}}=\frac{v_{1}}{\rho^{2}}e^{-\frac{\beta}{2}\rho^{2}}\left[1-\frac{1}{\sqrt{\beta}\rho}e^{-\frac{\beta\rho^{2}}{4}}\mathrm{erf}\left(\frac{\sqrt{\beta}\rho}{2}\right)\right]dv_{0}dv_{1}\]

This is plotted in the figure.

\includegraphics{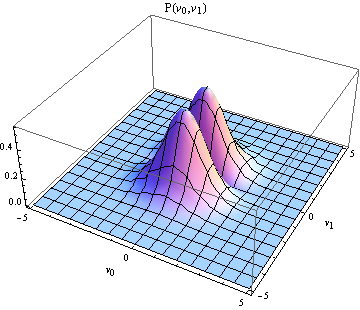}

When $a\neq0,$we still have an explicit solution

\[
P(v_{0},v_{1})\frac{dv_{0}dv_{1}}{v_{1}}=e^{-\frac{\beta}{2}\rho^{2}}\left[\rho^{\frac{1+\sqrt{1+4\beta_{1}}}{2}}F_{1}^{1}\left(\frac{1+\sqrt{1+4\beta_{1}}}{4},1+\frac{1}{2}\sqrt{1+4\beta_{1}},\frac{\beta r^{2}}{2}\right)\right]e^{-a\frac{v_{0}}{\rho}}\frac{1}{\rho}S_{11}\left(a,\frac{v_{0}}{\rho}\right)dv_{0}dv_{1}\]

which has a similar shape. It is easy to check that this is normalizable.

\section{Conclusion}

There is a large literature on Brownian motion in curved manifolds.
Mostly, this amounts to a study of the heat kernel of the Laplace-Beltrami
operator. This is justified in the over-damped limit, where velocity
(rather than acceleration) is proportional to the force. Our work
shows that there are unexpected subtleties in the more general case.
The velocities do not tend to the expected Maxwell-Boltzmann distribution
asymptotically, although for short times they might appear to do so.

For fluid mechanics, the phase space is infinite dimensional. It is
possible to embed the phase space into an infinite dimensional analogue
of the upper half plane (the space of complex symmetric matrices with
positive imaginary part), which has a very similar behavior for the
velocities. We plan to return to this case, which requires much deeper
mathematics than used in this paper, in a future publication. It is
hoped that the equilibrium velocity distribution of fluids under random
forces is experimentally accessible and gives useful some information
about strongly turbulent flows.

\section{Appendix: Formulas For Change Of Variables}

\[
v_{0}=-\rho\tanh\theta,\ v_{1}=\rho\epsilon\mathrm{sech}\theta\]

\[
dv_{0}=-d\rho\tanh\theta-\rho\mathrm{sech}^{2}\theta d\theta\]

\[
dv_{1}=\epsilon d\rho\mathrm{sech}\theta-\rho\epsilon\mathrm{sech}\theta\tanh\theta d\theta\]

\[
dv_{0}^{2}+dv_{1}^{2}=d\rho^{2}+\rho^{2}\mathrm{sech}^{2}\theta d\theta^{2}\]

\[
g_{ab}=\left[\begin{array}{cc}
1 & 0\\
0 & \rho^{2}\mathrm{sech}^{2}\theta\end{array}\right]\]

\[
g^{ab}=\left[\begin{array}{cc}
1 & 0\\
0 & \frac{1}{\rho^{2}}\cosh^{2}\theta\end{array}\right]\]

\[
dv_{0}\wedge dv_{1}=\rho\epsilon\mathrm{sech}\theta d\rho\wedge d\theta\]

\[
\mu=\frac{1}{\rho\mathrm{sech}\ \theta}\]

\[
\mu dv_{0}\wedge dv_{1}=d\rho\wedge d\theta\]

Thus these co-ordinates are canonically conjugate:

\[
\left\{ \rho,\theta\right\} =1\]

\section{Acknowledgement}

This work was supported in part by a grant from the US Department
of Energy under contract DE-FG02-91ER40685.


\begin{thebibliography}{10}
\bibitem{Arnold} V. I. Arnold, Ann. Inst. Poly. Genoble 16 (1966)
319.

\bibitem{ChandraStochastic} S. Chandrasekhar, Rev. Mod. Phys. 15
(1943) 1.

\bibitem{Kramers}H. A. Kramers, Physica,7 (1940) 284.

\bibitem{Kraichnan} R. H. Kraichnan, \emph{The closure problem of
turbulence theory}, Proc. Symp. Applied Mat., 13 (1965) 199.

\bibitem{SLE} R. M. Friedrich, \emph{The Global Geometry of Stochastic
Lœwner Evolutions, }arXiv:0906.5328v1 {[}math-ph{]}.

\bibitem{RigidBody} G. W. Ford, J. T. Lewis and J., McConnel, Phys.
rev. A19,(1979) 907; M. Efroimsky, J.Math.Phys. 41 (2000) 1854. arxiv:astro-ph/9909220

\bibitem{Chavel}~I. Chavel, \emph{Riemannian Geometry: A Modern
Introduction}, Cambridge University Press (2006)

\bibitem{Milnor} J. Milnor, Adv. Math. 21 (1976) 293.

\bibitem{hydrochennai} S. G. Rajeev, \emph{Geometry of the Motion
of Ideal Fluids and Rigid Bodies,}arXiv:0906.0184v1 {[}math-ph{]}.

\bibitem{AbrSt}M. Abramowitz and I. A. Stegun, \emph{Handbook of
Mathematical Functions,} p. 753, Dover (1972).
\end{thebibliography}
\end{document}